\pdfoutput=1
\documentclass[sn-nature,pdflatex]{sn-jnl}

\usepackage{graphicx}
\usepackage{booktabs,array}
\newcolumntype{L}[1]{>{\raggedright\arraybackslash}p{#1}}
\usepackage{amsmath,amssymb}
\usepackage{xcolor}
\usepackage{url}
\usepackage{etoolbox}
\makeatletter
\patchcmd{\@maketitle}{Corresponding author(s). E-mail(s):}{Corresponding author. E-mail:}{}{\PackageError{main}{Failed to patch the corresponding-author label}{}}
\makeatother
\graphicspath{{figures/}}

\begin{document}

\title[ForgeSystem]{ForgeSystem: An Architecture for Demand-Specific Artefact Production with AI Agents}

\author[1,2]{\fnm{Qingfeng} \sur{He}}
\author[1,2]{\fnm{Zhui} \sur{Zhu}}
\author[1,2]{\fnm{Yaojian} \sur{Chen}}
\author[1,3]{\fnm{Shangzhan} \sur{Li}}
\author[1,2]{\fnm{Leshan} \sur{Li}}
\author[1]{\fnm{Tingjian} \sur{Zhang}}
\author[1,4]{\fnm{Qian} \sur{Xie}}
\author*[1,2]{\fnm{Yuxuan} \sur{Li}}\email{liyuxuan@modelbest.cn}
\author[1,2]{\fnm{Xu} \sur{Han}}
\author[1,2]{\fnm{Zhiyuan} \sur{Liu}}
\affil[1]{\orgname{ModelBest Inc.}, \orgaddress{\city{Beijing}, \country{China}}}
\affil[2]{\orgdiv{Department of Computer Science and Technology}, \orgname{Tsinghua University}, \orgaddress{\city{Beijing}, \country{China}}}
\affil[3]{\orgname{Harbin Institute of Technology}, \orgaddress{\city{Harbin}, \country{China}}}
\affil[4]{\orgdiv{Columbia Business School}, \orgname{Columbia University}, \orgaddress{\city{New York}, \country{USA}}}

\abstract{General-purpose software serves many different demands, but this flexibility costs performance and efficiency on any specific demand. Software tailored to a specific demand avoids these losses, yet developing it requires substantial time and effort from human experts. AI agents offer an alternative: they can design, implement and test artefacts tailored to individual demands. We call this production paradigm Forge. Forging for a demand is economically justified when the cumulative benefit of specialisation exceeds the total cost of construction, verification and maintenance. To make Forge a repeatable production process, we propose ForgeSystem, which consists of a control layer, a forge loop and a recursive self-improvement (RSI) loop. The control layer selects which artefact serves each demand and decides which demands to forge, when to forge them and from which starting point; the forge loop iterates between production and verification until a candidate artefact passes verification; and the RSI loop draws on accumulated production records to supply starting points, refine the harness and train models. We implement ForgeSystem for training frameworks, decoding megakernels and individual computational kernels. Forged training engines reach $1.047$--$1.332\times$ the model-FLOPs utilisation of their reference frameworks, forged megakernels achieve a geometric-mean decoding speedup of $1.21\times$ over SGLang across thirteen configurations, and forged kernels match or exceed expert libraries in eighteen of twenty-two configurations, while satisfying task-specific quality requirements. Controlled comparisons show that explicit quality requirements steer agents towards optimisations that preserve training quality. We also discuss a gradual transition to Forge through service routing, and how demand-specific production may shift value accumulation and trust towards production systems.}


\maketitle

\section{Introduction}

General-purpose software serves diverse demands through a shared implementation, but the design compromises needed to accommodate different scenarios limit its performance and efficiency for specific demands. Specialised software, by contrast, is designed around a specific demand and can exploit its particular conditions for more thorough optimisation. For example, an inference engine tailored to a particular model, hardware platform and workload can exploit the interactions among model structure, hardware characteristics and workload patterns to optimise overall execution efficiency. Realising these benefits has traditionally required experts to make demand-specific design choices, implement them and conduct experiments. Repeating this work for each new demand makes specialisation costly and limits the range of demands it can economically serve.

AI agents are beginning to undertake this construction work. VibeServe generates inference systems tailored to specified models, hardware and workloads\cite{vibeserve2026}. Bespoke OLAP constructs analytical engines from scratch for given query workloads\cite{bespokeolap2026}. Similar agents have been built for video-inference acceleration and kernel generation (Appendix~\ref{app:other_systems})\cite{sol2026,cake2026,ksearch2026}. Despite differences in their outputs, these systems share a production structure: a concrete demand specifies the objectives and constraints, and an agent organises design, implementation and evaluation to produce a specialised artefact.

We call this production paradigm \emph{Forge}: agents organise resources and iteratively construct, evaluate and refine specialised artefacts for explicit demands. The demand determines the scope of construction, from an individual computational kernel to a software system comprising interacting components. Forge reuses construction capabilities across demands to produce artefacts tailored to each demand. Its economic rationale is that the cumulative benefit of specialisation can exceed the total cost of construction, verification and maintenance.

Making Forge a repeatable production process requires organising long-horizon construction, maintaining quality throughout optimisation and using accumulated experience to improve subsequent production. To meet these requirements, we propose ForgeSystem, which consists of a control layer, a forge loop and a recursive self-improvement (RSI) loop (Fig.~\ref{fig:system}). The control layer determines which artefact serves each demand and decides which demands to forge, when to forge them and from which starting point. The forge loop constructs and verifies artefacts, keeping production separate from verification. Production combines deep specialisation, vertical integration and massive exploration. Verification consists of process management, which constrains how agents carry out construction, and artefact judgement, which checks the artefacts they produce. The RSI loop stores each delivered artefact together with its demand and construction trace, and draws on these records to improve subsequent forges by providing starting points, refining the harness and training the underlying model.

We implement ForgeSystem at three levels---complete training frameworks, decoding megakernels and individual computational kernels---yielding ForgeTrain, ForgeMegakernel and ForgeKernel\cite{forgetrain2026,forgemegakernel2026,forgekernel2026}. ForgeTrain's model FLOPs utilisation is $1.047$--$1.332\times$ that of the reference frameworks\cite{forgetrain2026}, and ForgeMegakernel achieves a geometric-mean speedup of $1.21\times$ over SGLang across thirteen configurations\cite{forgemegakernel2026}. ForgeKernel achieves a speedup of at least $1.0\times$ over the references in eighteen of twenty-two expert-library configurations and all seven configurations evaluated against kernel-generation systems\cite{forgekernel2026}. Quality evaluations show comparable training outcomes and inference quality, together with correct operator outputs. Controlled comparisons in training-framework construction further show that explicit quality requirements can guide agents towards optimisations that preserve training quality\cite{forgetrain2026}.

Our contributions are fourfold:
\begin{enumerate}
\item We formulate Forge as a production paradigm in which agents construct specialised artefacts for explicit demands, and articulate an economic criterion based on the benefits of specialisation and the total cost of production.
\item We propose ForgeSystem, which consists of a control layer, a forge loop and an RSI loop, instantiate it in training frameworks, megakernels and computational kernels, and evaluate its performance and correctness through experiments, ablations and controlled comparisons.
\item We outline a gradual transition path from the general-purpose paradigm to the Forge paradigm, in which general-purpose and forged artefacts coexist and service progressively shifts to forged artefacts that pass verification.
\item We analyse two potential long-term implications of demand-specific production: a shift in value accumulation towards production systems, and an extension of trust from individual artefacts to the systems that produce them.
\end{enumerate}

\section{Background}

The Forge paradigm introduced above builds on the practice of specialisation and the evolution of model technology. This section compares three production modes for differentiated demands and discusses how changes in model capability and invocation cost affect the feasibility and economic scope of demand-specific construction.

\subsection{Three production modes for differentiated demands}

For many different demands, software production either reuses one implementation across demands or redesigns around an individual demand. The first consolidates the shared parts of demands into a \emph{general-purpose artefact}; the second, carried out by human experts or by AI agents, makes the artefact serve the goals and environment of one demand.

\textbf{General-purpose artefacts.} Megatron-LM supports multiple models and training configurations through a unified framework, parallelism abstractions and optimisation components\cite{shoeybi2019megatron,narayanan2021megatron}. Specific demands are accommodated by selecting configurations and combining components within this framework, with optimisation constrained by existing interfaces and execution strategies. Design compromises retained to accommodate different models, hardware platforms and workloads make it difficult for a shared implementation to realise the full performance and efficiency potential of every configuration. Software product lines and mass customisation describe this production approach based on a predefined space of variation\cite{clements2001spl,pine1993mass}: demands within that space can be met through configuration, whereas specialisation beyond it requires additional development.

\textbf{Human customisation.} DeepSeek's models and training systems exemplify expert teams co-designing architecture, communication and computational kernels around a concrete demand\cite{deepseekv2,deepseekv3,fireflyer2024}. Their infrastructure report shows that on their self-built cluster of ten thousand PCIe A100 GPUs, the in-house HFReduce communication library reaches inter-node allreduce bandwidths of 6.3--8.1 GB/s where the general-purpose NCCL reaches only 1.6--4.8 GB/s, and the in-house training framework roughly halves data-parallel and FSDP training time relative to the corresponding PyTorch implementations\cite{fireflyer2024}. These results show that the compromises a general-purpose implementation retains to accommodate different hardware platforms and workloads leave substantial performance unexploited in a concrete configuration, and that customising around the demand recovers it. Human customisation extends specialisation across layers, but still depends on experts to choose solutions, implement them and run experiments. Its ability to cover more demands depends on whether these activities can be repeated at an acceptable cost.

\textbf{AI customisation.} In AI customisation, agents organise resources, search for solutions, implement designs and iterate through experiments around specific demands. ForgeTrain is one example: an agent constructs a specialised training framework for a specified model and hardware configuration\cite{forgetrain2026}.

The shift from human to AI customisation ties the feasibility of demand-specific production closely to agents' construction capabilities and the costs of production. Agents must be able to construct artefacts that meet the demand, and the benefits of specialised artefacts must cover the investment in construction, verification and maintenance. Improving model capability and declining invocation costs are changing both conditions.

\subsection{Why now?}

Forging offers an economic advantage when the benefits of specialisation exceed the total cost of production. For a demand $d$ and a forged artefact $a$ that satisfies its quality requirements, this condition is
\begin{equation}
B_d(a) > C_d(a),
\label{eq:criterion}
\end{equation}
where $B_d(a)$ is the cumulative benefit of $a$ over the artefact that would serve $d$ without this forge while the demand remains in service, and $C_d(a)$ is the total cost of producing and maintaining $a$, including construction, the execution of verification, necessary human input and maintenance over that period. Both quantities are measured on a common basis, such as compute expenditure at equal workload and quality. This criterion connects the value of demand-specific optimisation to the investment required to realise it. Because $a$ is unknown before construction, the criterion is applied in expectation over the artefacts that forging from a given starting point may produce, $\mathbb{E}[B_d(a)] > \mathbb{E}[C_d(a)]$. The condition fails when specialisation brings little benefit or when the full cost of experimentation, verification and integration remains high; Appendix~\ref{app:scope} discusses these limits.

\begin{figure}[!htbp]
\centering
\includegraphics[width=\textwidth]{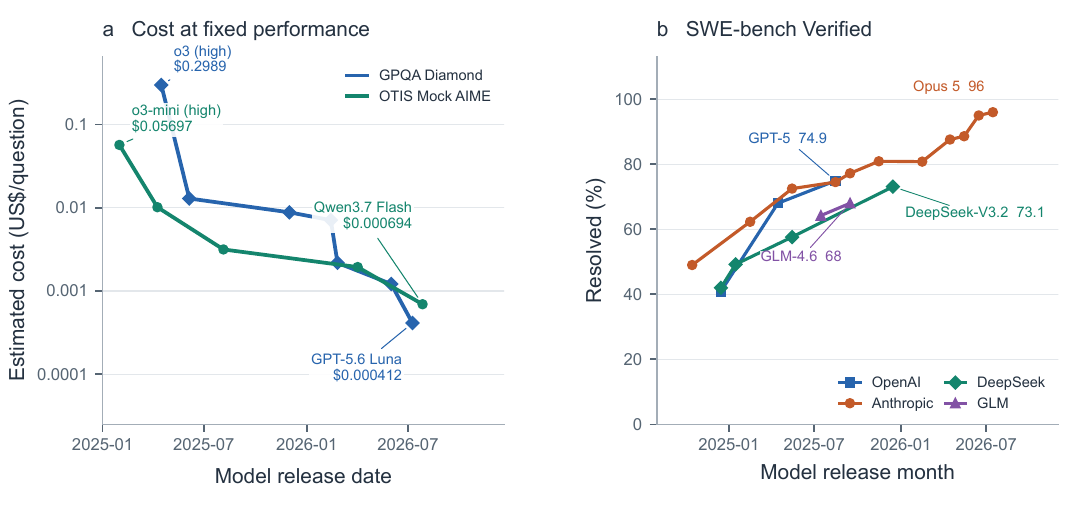}
\caption{\textbf{Model capability and task cost at fixed performance.} a, Estimated cost per question at fixed performance on GPQA Diamond and OTIS Mock AIME. Cost data are from Epoch AI\cite{epochthought2026,roodman2026inferencecost,epochgpqa,epochotisaime}. b, Vendor-reported SWE-bench Verified\cite{jimenez2024swebench,swebenchverifieddocs2024} resolution rates, with tools and evaluation settings varying across records. Scores are from system cards, release announcements and technical reports by OpenAI\cite{openaio12024,openaio32025,openaigpt52025}, Anthropic\cite{anthropic2024swe,anthropic2025sonnet37,anthropic2025claude4,anthropic2025opus41,anthropic2025sonnet45,anthropic2025opus45,anthropic2026opus46,anthropic2026opus47,anthropic2026opus48,anthropic2026fable5,anthropic2026opus5}, DeepSeek\cite{deepseekv3,deepseekr1,deepseekr10528,deepseekv32} and Zhipu AI\cite{glm452025,glm462025}.}
\label{fig:curves}
\end{figure}

Improving model capability and declining invocation costs are making this condition attainable for a broader range of demands. Scaling training investment and increasing model capability density provide complementary routes towards more capable and economical models\cite{kaplan2020scaling,hoffmann2022chinchilla,xiao2024densing}. At the application level, reported SWE-bench Verified results show advances in agents' ability to resolve real repository issues, expanding the implementation and revision work they can undertake (Fig.~\ref{fig:curves}b)\cite{jimenez2024swebench,swebenchverifieddocs2024}. At fixed performance thresholds, task-cost estimates for GPQA Diamond and OTIS Mock AIME show how later models can achieve comparable outcomes at lower invocation cost (Fig.~\ref{fig:curves}a)\cite{epochthought2026,epochgpqa,epochotisaime}. Stronger capabilities bring more specialised constructions within reach, while cheaper model calls reduce the expenditure associated with repeated generation and refinement. Together, these developments improve the technical and economic conditions for forging. The next section introduces ForgeSystem, which translates these conditions into a repeatable process of production, verification and improvement.

\section{ForgeSystem}

\begin{figure}[!htbp]
\centering
\includegraphics[width=\textwidth]{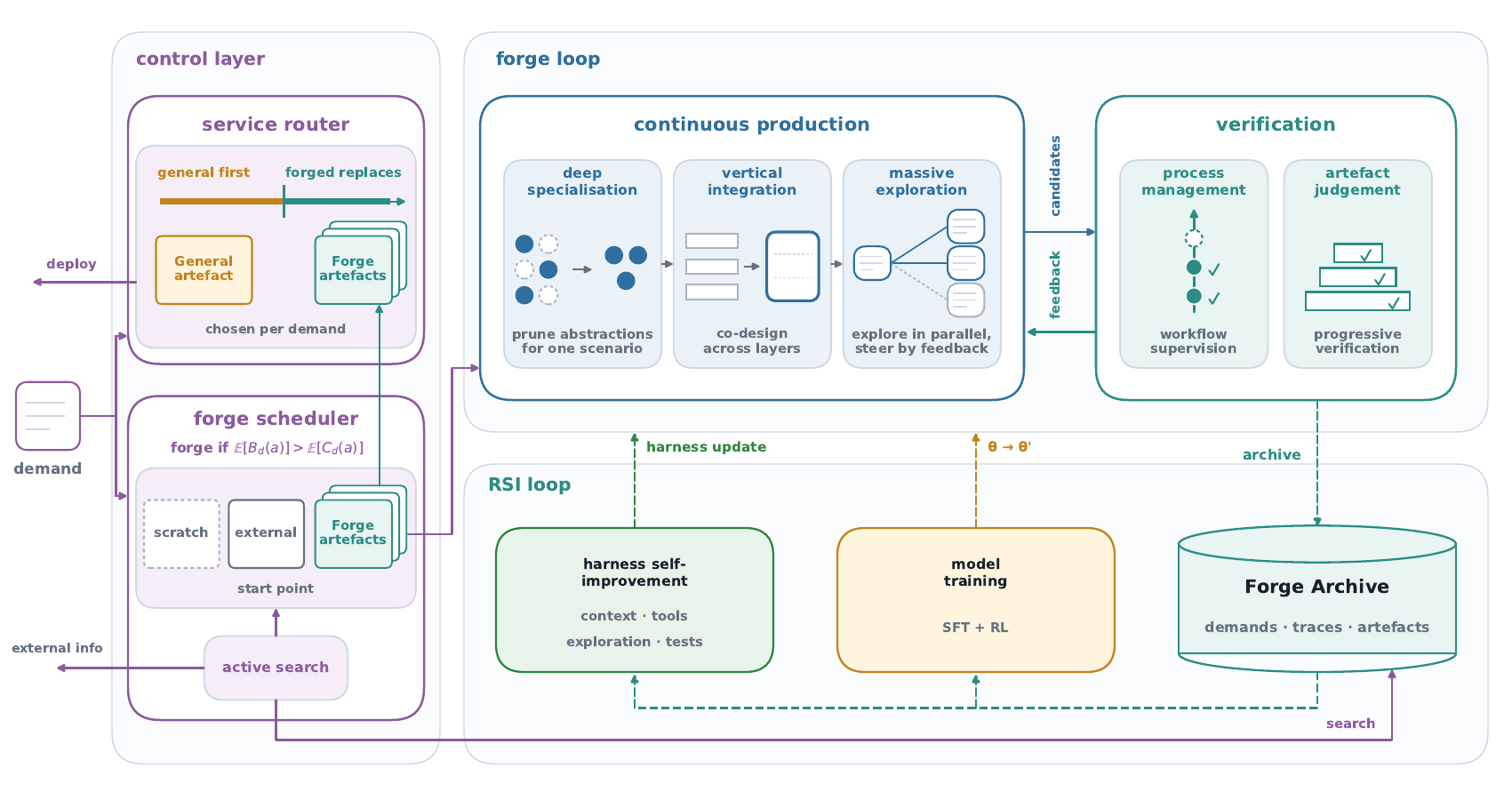}
\caption{\textbf{ForgeSystem.} Left: the control layer. The service router determines which artefact serves each demand; the forge scheduler uses active search to gather information from external sources and the Forge Archive, chooses among three kinds of starting point and launches a forge when Eq.~(\ref{eq:criterion}) holds in expectation. Upper right: the forge loop, where production combines deep specialisation, vertical integration and massive exploration, and verification combines process management with artefact judgement. Lower right: the RSI loop, in which the Forge Archive stores each delivered artefact with its demand and construction trace and supplies data for forged-artefact feedback, harness self-improvement and model training. Solid arrows: demands, deployment, forge tasks, artefact flow, judgement feedback and searches; dashed arrows: archiving, archive data and updates across forges.}
\label{fig:system}
\end{figure}

ForgeSystem consists of a control layer, a forge loop and a recursive self-improvement (RSI) loop (Fig.~\ref{fig:system}). The control layer comprises a service router and a forge scheduler: the service router determines whether the general-purpose artefact or a forged artefact serves each demand, and the forge scheduler decides which demands to forge, when to forge them and from which starting point (Section~\ref{sec:scheduler}). The forge loop constructs and verifies artefacts, keeping production separate from verification (Section~\ref{sec:loop}). The RSI loop stores each delivered artefact in the Forge Archive together with its demand and construction trace, and improves subsequent forges through forged-artefact feedback, harness self-improvement and model training (Section~\ref{sec:rsi}).

\subsection{Forge scheduler}
\label{sec:scheduler}

The forge scheduler decides which demands to forge, when to forge them and from which starting point. To make these decisions, it continuously gathers two kinds of information through active search. External information covers newly released general-purpose and specialised artefacts, as well as advances in models and hardware; these changes alter the benefit that forging can bring to each demand. Internal information consists of the forged artefacts and construction traces of similar demands in the Forge Archive; it indicates which artefacts the system can already construct and at what cost. On the basis of this information, the scheduler judges which demands are worth forging. A forge can begin from scratch, from an external implementation retrieved by active search, or from a forged artefact in the Forge Archive. These three kinds of starting point differ in how much of the demand they already meet and how much construction remains, so the expected benefit and cost of forging vary with the starting point. The scheduler selects the most suitable of the available starting points and launches a forge when Eq.~(\ref{eq:criterion}) holds in expectation.

\subsection{Forge loop}
\label{sec:loop}

The forge loop starts from the demand and starting point selected by the forge scheduler and iterates between production and verification: the production side generates and improves candidate artefacts, and the verification side checks them and returns feedback, until a candidate passes verification and is delivered. The production side may adjust its construction in response to this feedback, but must not lower verification requirements or modify trusted references.

\subsubsection{Production}

\noindent\textbf{Deep specialisation.} Forging decouples scenarios and produces a specialised artefact for each. Specifying the scenario allows production to reduce general-purpose abstraction layers, interfaces and optional mechanisms, removing unnecessary branches and conversions. Differences otherwise handled at runtime become known conditions for specialising implementations during construction.

\noindent\textbf{Vertical integration.} Reducing abstraction layers enables joint resource organisation across boundaries retained for compatibility with other scenarios. A megakernel, for example, coordinates GPU computation, data loading, on-chip storage and synchronisation to reduce intermediate transfers; register usage, storage capacity and parallelism constrain its benefits\cite{forgemegakernel2026}. The demand and measured outcomes determine the scope of integration, whether across layers or between simple components.

\noindent\textbf{Massive exploration.} Resources, their combinations and integration choices span a large design space, of which human experts can examine only a small fraction. Agents propose hypotheses, construct candidates and run experiments in parallel, allocating further exploration according to judgement feedback. A branch may require several successive changes before its benefits emerge. More attempts create more opportunities for improvement, while accumulated knowledge, experience and feedback reduce redundant trials.

\subsubsection{Verification}

\noindent\textbf{Process management.} Process management organises the agent's task decomposition, scheme exploration, implementation changes and experiment planning during forging, while constraining the contexts, tools and resources it uses. The system continuously stores the demand conditions, construction state, generated artefacts, execution trace, verification results and dependencies of each forging step, so that every step can be re-examined, recovered and reused. Verification feedback during forging is used to adjust the search direction, allocate exploration resources and decide whether to continue modifying; a state without sufficient evidence cannot be treated as a construction basis and propagated directly to subsequent steps. Process management thereby reduces unsupported modifications and error accumulation, enabling long-horizon agents to maintain traceability and controllability during continued exploration.

\noindent\textbf{Artefact judgement.} The system treats components, interfaces and intermediate artefacts checked under explicit conditions as a trusted foundation, and treats each addition or modification as an untrusted sub-operation. For an untrusted sub-operation, it first uses checks that can expose problems at relatively low cost, promptly eliminating candidates with compilation failures, interface inconsistencies, obvious errors or mismatches with the reference; candidates that pass these checks then undergo verification covering broader properties and conditions. Verification first establishes the artefact's correctness and task quality, and then evaluates performance and resource cost subject to these quality constraints. Checks differ in cost, scope and reliability, and each provides evidence only about the properties and conditions it covers; passing a local check does not directly establish that the complete artefact is acceptable. When the available evidence is insufficient to support a judgement, more comprehensive or reliable checks are required to form the evidence needed for delivery.

\subsection{Recursive self-improvement}
\label{sec:rsi}

\noindent\textbf{Forged-artefact feedback.} Delivered forged artefacts and their components can serve as starting points for subsequent construction. The Forge Archive stores each delivered artefact together with its demand and construction trace, and the forge scheduler can select a starting point from it (Section~\ref{sec:scheduler}). Depending on the differences between the selected artefact and the demand, subsequent construction can directly reuse, adapt or reconstruct it, or combine components from several artefacts. When the archive already holds a forged artefact that meets the demand, the forge scheduler still estimates the expected benefit and cost of forging again, with the benefit measured against this forged artefact; if the expected benefit exceeds the expected cost (Eq.~(\ref{eq:criterion})), it forges again with this forged artefact as the starting point. Such reuse changes how exploration is organised: solved parts need not be rebuilt, new attempts can focus on the unmet parts of the demand, and existing components can be recombined.

\noindent\textbf{Harness self-improvement.} We incorporate ontology engineering into the harness, organising domain concepts, component relations, dependencies and applicability conditions into a retrievable knowledge graph that provides agents with the domain knowledge needed for construction. Execution feedback guides changes to context provision, tool use, the organisation of exploration, and testing. After the harness is updated, the system compares artefact quality and production cost on subsequent forging tasks to assess whether these changes improve the production process.

\noindent\textbf{Self-improvement through training.} Because archived delivered artefacts and their construction traces have already passed independent judgement, they provide quality-screened examples for supervised learning of implementation choices, cross-component design and search. Archived demands can also serve as reinforcement-learning tasks, with rewards assigned by the verification side to newly generated rollouts. Training requires long-term investment and validation to internalise experience. Updated models return to forging, closing an RSI loop in the narrow sense: judgement-screened construction traces train the model that produces subsequent artefacts, with gains assessed through subsequent artefact quality, construction cost and search efficiency.

\section{Empirical evaluation of ForgeSystem}
\label{sec:evaluation}

We implement ForgeSystem (Fig.~\ref{fig:system}) in three settings: training frameworks, decoding megakernels and individual computational kernels. The corresponding systems are ForgeTrain, ForgeMegakernel and ForgeKernel\cite{forgetrain2026,forgemegakernel2026,forgekernel2026}. We examine whether the specialised artefacts produced by these systems deliver performance gains, satisfy their quality requirements and provide starting points for subsequent production, and what construction effort they require.

\subsection{Experimental setup}

Hardware and software stacks depend on the artefact. ForgeTrain forges training frameworks in seven model--hardware settings: Qwen3 0.6B\cite{yang2025qwen3}, MiniCPM4 0.5B, MiniCPM4 8B\cite{minicpm4}, MiniCPM5 1B\cite{minicpm5repo} and MiniCPM5 16B-A3B on NVIDIA H100, with Megatron-LM 0.15 as the reference, and MiniCPM5 1B and MiniCPM5 130M on Huawei Ascend 910, with MindSpeed\cite{mindspeed} (Megatron core 0.12.1) and PyTorch\cite{paszke2019pytorch} (torch\_npu) as the respective references. The H100 experiments use PyTorch 2.8.0 with CUDA 12.9. The sequence length is fixed at 4096. Throughput is measured by model FLOPs utilisation (MFU): exact operator-level FLOP counts give achieved FLOPs per second, which is divided by peak compute throughput times the device count; the forged engine and reference use the same accounting\cite{forgetrain2026}.

ForgeMegakernel forges megakernels on an H100 80GB GPU with 3350\,GB/s peak HBM3 bandwidth. References include tuned SGLang 0.5.18\cite{zheng2024sglang}, vLLM 0.27.1\cite{kwon2023vllm}, llama.cpp\cite{llamacpp} and the MPK megakernel compiler\cite{cheng2025mpk}. The metric, model bandwidth utilisation (MBU), is the bytes that a decode step must move divided by the bytes transferable at peak bandwidth over the same time. Under identical model, batch size, sequence length, byte accounting and peak bandwidth, the candidate-to-reference MBU ratio equals the reference-to-candidate decode-latency ratio\cite{forgemegakernel2026}.

ForgeKernel forges individual computational kernels on Hopper H100 and Blackwell B200 GPUs, using reference kernels from expert libraries including FlashAttention-4\cite{zadouri2026flashattention4}, DeepGEMM\cite{deepgemm2025}, cuBLAS\cite{cublas}, cuDNN\cite{chetlur2014cudnn}, FlashInfer\cite{ye2025flashinfer}, FlashMLA\cite{flashmla2025} and FlashQLA\cite{flashqla2026}, together with kernels released by other generation systems. Device execution time is measured under cold-L2 conditions with the same timing protocol for candidate and reference; the speedup is the ratio of reference to candidate execution time\cite{forgekernel2026}.

\subsection{Performance, quality and cost of specialised artefacts}

\begin{figure}[!htbp]
\centering
\includegraphics[width=\textwidth]{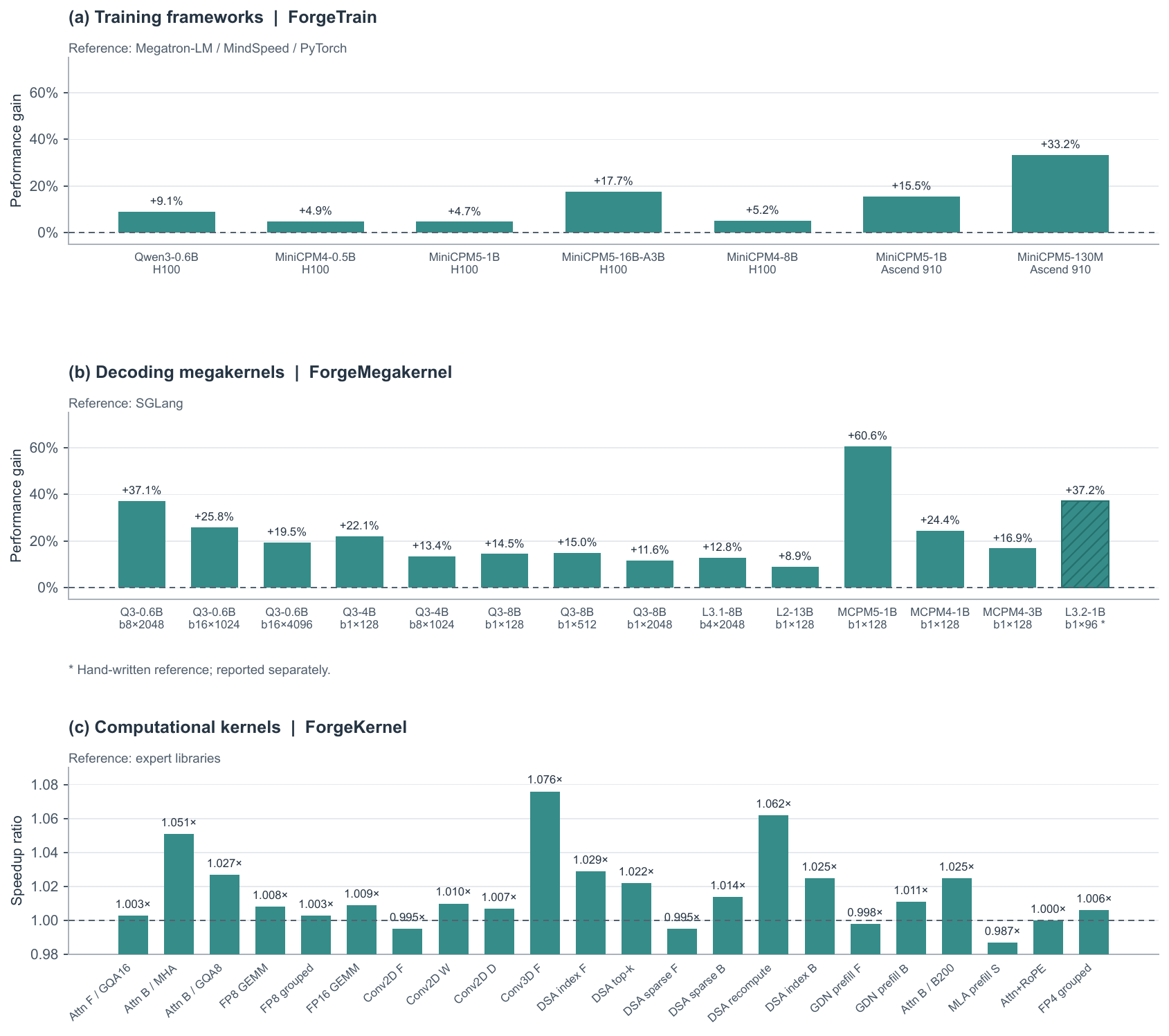}
\caption{\textbf{Performance of specialised artefacts.} a, ForgeTrain's MFU gains over its reference framework in seven model--hardware settings. b, ForgeMegakernel's per-configuration decoding gains over SGLang. c, ForgeKernel's speedup ratios in twenty-two expert-library configurations, with the dashed line at $1.0\times$ marking reference performance. Panels a and b express gains as $(r-1)\times100\%$, where $r$ is the performance ratio of the forged artefact to its reference; panel c directly shows $r$, the ratio of reference to candidate execution time. ForgeTrain uses the relative MFU gains in the source paper's Table 2 rather than recomputing them from rounded values; ForgeMegakernel uses the MBU data of the source paper's Fig.~3; ForgeKernel uses the per-configuration expert-library results in the source paper, excluding generation-system reference configurations; each value is the best generation-0 optimisation run for its configuration and excludes gains from re-forging\cite{forgetrain2026,forgemegakernel2026,forgekernel2026}. Q3, L and MCPM denote Qwen3\cite{yang2025qwen3}, Llama\cite{touvron2023llama2,grattafiori2024llama3} and MiniCPM\cite{hu2024minicpm,minicpm4}, $b\times S$ denotes batch size and sequence length, and F/B/W/D denote forward, backward, weight gradient and input gradient.}
\label{fig:performance}
\end{figure}

\textbf{Performance of specialised artefacts.} We first test whether forged artefacts outperform mature frameworks for their specified demands. In seven model--hardware settings, ForgeTrain's MFU is $1.047$--$1.332\times$ that of the corresponding reference framework. Dense models on H100 reach approximately $1.05$--$1.09\times$, MiniCPM5 16B-A3B reaches $1.177\times$, and the two Ascend 910 settings reach $1.155\times$ and $1.332\times$, respectively\cite{forgetrain2026}. These results demonstrate the performance headroom available when a training system is constructed for a particular model and hardware configuration (Fig.~\ref{fig:performance}a).

For decoding, ForgeMegakernel constructs megakernels for fourteen configurations. Thirteen configurations use SGLang as a common reference; they achieve 50.5\%--85.9\% MBU and a geometric-mean speedup of 1.21$\times$ over SGLang. On the same configurations, the forged megakernels achieve a geometric-mean speedup of 1.54$\times$ over MPK. The fourteenth configuration, Llama-3.2-1B, is forged against a reproduced hand-written megakernel and is therefore excluded from the aggregates above. The forged kernel reaches 74.0\% MBU, comparable to the hand-written reference's 74.6\%, with speedups of 1.37$\times$, 1.50$\times$, 1.52$\times$ and 1.84$\times$ over SGLang, vLLM, MPK and llama.cpp; its gain over SGLang is marked separately in Fig.~\ref{fig:performance}b\cite{forgemegakernel2026}.

ForgeKernel has two reference groups. In twenty-two configurations against pinned expert libraries, eighteen have a relative speedup of at least $1.0\times$, a further three are within 1\% of the reference, and the remaining configuration (sparse MLA prefill on B200) trails its reference by 1.3\%; the geometric-mean speedup is 1.016$\times$. In seven configurations against published kernels from K-Search\cite{ksearch2026}, Kernel Design Agents\cite{kda2026} and CAKE\cite{cake2026}, every result has a relative speedup of at least $1.0\times$; the geometric mean is 1.394$\times$ and the maximum is 2.706$\times$\cite{forgekernel2026}. Per-configuration results for the twenty-two expert-library configurations are shown in Fig.~\ref{fig:performance}c.

\textbf{Correctness and quality.} Performance benefits must satisfy the quality constraints in the demand. ForgeTrain extends validation to long-run training and downstream evaluation in three settings: MiniCPM4 0.5B on H100, and MiniCPM5 1B and MiniCPM5 130M on Ascend. For MiniCPM5 1B, after the same 48,000-step decay training, the forged engine and MindSpeed reach final losses that differ by only 0.0001 (the two values are 1.3227 and 1.3226). Across the three comparisons, the average downstream score differences (forged minus reference) are $+0.68$, $-0.68$ and $+0.93$ points, each within one point. These results establish training-quality parity between the forged engines and their references in the tested settings\cite{forgetrain2026}.

ForgeMegakernel uses a float64 forward computation from the same checkpoint as an independent reference to check numerical error under identical prompts and teacher-forced input sequences, and then evaluates the kernels in SGLang. On 200 GSM8K\cite{cobbe2021gsm8k} test questions, accuracy differs from the baseline by at most 1.5 percentage points for every configuration, with seven configurations matching or exceeding the baseline. Latency per decoded token in serving is lower than the baseline in every configuration, with a geometric-mean speedup of 1.11$\times$\cite{forgemegakernel2026}. This result and the 1.21$\times$ isolated decode-step speedup use different measurement boundaries, reflecting kernel performance and the benefit after service integration, respectively.

Before timing, ForgeKernel requires each candidate to pass the official or benchmark tolerance test associated with its operator source; only candidates that pass these checks enter performance measurement\cite{forgekernel2026}. The objects of verification therefore differ: training frameworks require checks of training behaviour and final model quality; decode implementations require numerical and serving-quality checks; and individual operators must satisfy output-correctness requirements. Performance results and these quality records jointly support acceptance; the ablations in Appendix~\ref{app:ablations} examine the role of the constraints.

\textbf{Forged-artefact feedback.} ForgeKernel instantiates the re-forging path of Section~\ref{sec:rsi}: an accepted kernel enters subsequent runs as the performance reference that new candidates must surpass, and construction is organised around analysing and beating it. Re-forging the same demand from this starting point continues to yield gains: GQA attention forward improves by 9.9\% over its accepted predecessor in one further run and by another 4.2\% in the next, reaching about 15\% above the expert library; DSA (DeepSeek sparse attention) sparse-attention forward improves by 7.2\% with its own previous output as the reference\cite{forgekernel2026}. These results show that an accepted artefact can provide a performance starting point for subsequent production, and that forging again for the same demand can still yield further gains.

\textbf{Construction effort.} ForgeTrain typically needs about 12--20 rounds on H100 and 3--5 rounds on Ascend to first exceed its reference\cite{forgetrain2026}. ForgeMegakernel records time, token usage and billing by round; median effort per kernel is 47 rounds, 14.7 agent-hours and US\$177, with billed costs ranging from US\$77 to US\$414\cite{forgemegakernel2026}. A one-time construction cost of hundreds of dollars can be amortised through compute savings from long-term, repeated deployment. For a continuously running inference service, these benefits accumulate with request volume.

\section{Discussion}
\label{sec:discussion}

Extending Forge to more domains involves not only producing artefacts for specific demands, but also the gradual adoption of this production approach and the sustained accumulation and trustworthy delivery that follow.

\subsection{A gradual transition to Forge}
\label{sec:general_to_forge}

The transition from the general-purpose paradigm to Forge can proceed through the coexistence of general-purpose and forged artefacts. A new demand may require immediate deployment, whereas constructing and verifying a specialised artefact takes time, so a general-purpose artefact can provide initial service. The service router in the control layer of Fig.~\ref{fig:system} supports this transition: while the general-purpose artefact serves the demand, the forge loop proceeds concurrently; once the forged artefact passes verification and meets deployment and migration requirements, the service router can switch subsequent work to it. A better artefact obtained by forging again later can replace the artefact in service in the same way. This follows a similar rationale to the Strangler Fig approach in software modernisation, which maintains service through the coexistence and gradual replacement of old and new implementations\cite{fowler2024strangler}. As forged artefacts enter service, users can adopt Forge progressively for individual demands without waiting for all demands to complete migration at once.

\subsection{Durable value shifts to production systems}

In the general-purpose-artefact era, the durable asset is the artefact: frameworks are maintained for decades, and their user bases constitute their value. Under Forge, an artefact is typically produced for a single demand and then deployed and maintained for as long as that demand persists. Upon delivery, the artefact is stored in the Forge Archive together with its demand and construction trace, and these records serve subsequent production as references, starting points and training data. What appreciates across demands is therefore not only an individual artefact but the production system itself: forged-artefact feedback supplies new construction starting points, harness self-improvement refines the production process, and model training incorporates accumulated production experience into subsequent forges. Amortisation shifts from the artefact to the accumulation behind it; whoever holds that accumulation is more likely to control production. General-purpose frameworks provide baselines for judging the gains and applicability of forged artefacts. The tendency for lower design costs to enable specialisation also echoes the shift towards domain-specific architectures\cite{hennessy2019golden}.

\subsection{Trust mechanisms extend to production systems}

Because each forged artefact is built for a specific demand, it may be generated for the first time, making it difficult to accumulate quality evidence through widespread use of the same artefact as general-purpose products do. The quality attributes of generated artefacts differ in how difficult they are to verify: some can be checked directly before adoption, some can only be judged through actual use, and others remain difficult for users to evaluate accurately even after use\cite{nelson1970experience,darby1973credence}. Demand-side users can still directly check properties that are inexpensive to verify. When quality appears only after long observation or cannot be independently evaluated by users, however, broader testing and longer observation also incur costs, so exchange cannot always depend on buyers directly checking every attribute\cite{barzel1982measurement}.

Where users cannot inspect every product themselves, trust does not rest on a single mechanism but on several layers of mechanisms. The first is the producer's brand and reputation: users identify a continuing producer through its brand and judge new outputs by its past deliveries; a single failure can damage an established reputation, and the prospect of lost future returns gives the producer an incentive to maintain quality\cite{wernerfelt1988umbrella,klein1981market}. The second is entry requirements: when users cannot distinguish quality, minimum quality standards and licensing keep low-quality supply out of the market and constrain the capabilities that producers must have\cite{leland1979licensing,shapiro1986licensing}. Medical care is a typical example: institutions do not judge each treatment on the patient's behalf, but set shared entry requirements and norms of conduct for services that are difficult to evaluate\cite{arrow1963uncertainty}. The third is standards and certification: standards specify which properties are evaluated, what evidence is required and which tasks and conditions are covered, and certification checks compliance with those standards within a defined scope.

Under the Forge paradigm, a system of trust should be built around production systems alongside the verification of individual artefacts. Brand and reputation accumulate the delivery record of a given production system, enabling users to choose based on its record on relevant tasks and making quality failures affect its subsequent adoption and returns. In application domains that require conditions for entry, entry requirements specify the basic conditions a production system must meet before supplying artefacts. Standards define requirements for quality control in the production process, artefact evaluation and evidence retention; certification checks whether a production system meets these requirements within a specified scope of tasks and conditions of use. Verification of the current artefact thus provides direct evidence, while the production system's reputation, eligibility to operate and certification results provide grounds for trust across artefacts, together supporting users' judgements of delivery quality.

\section{Conclusion}

We propose Forge: agents construct specialised artefacts for explicit demands when the benefits of specialisation exceed the total cost of production. To organise this paradigm as a repeatable production process, we introduce ForgeSystem, which consists of a control layer, a forge loop and an RSI loop. The control layer selects which artefact serves each demand and decides which demands to forge, when to forge them and from which starting point; the forge loop iterates between production and verification until a candidate artefact passes verification and is delivered; and the RSI loop draws on accumulated production records to improve subsequent forges through forged-artefact feedback, harness self-improvement and model training.

We instantiate this system in training frameworks, decoding megakernels and individual computational kernels. Experiments show that ForgeTrain's model FLOPs utilisation is $1.047$--$1.332\times$ that of the reference frameworks, ForgeMegakernel achieves a geometric-mean speedup of $1.21\times$ over SGLang across thirteen decoding configurations, and ForgeKernel achieves a speedup of at least $1.0\times$ over the references in eighteen of twenty-two expert-library configurations and all seven configurations evaluated against kernel-generation systems. Quality evaluations show that these performance gains meet the corresponding task-specific quality requirements; controlled comparisons in training-framework construction further show that explicit quality requirements steer agents towards optimisations that preserve training quality. These results show how agent-driven specialisation can extend beyond one-off generation to a production process combining production, verification and accumulated experience.

Service routing provides a path for a gradual transition from the general-purpose paradigm to the Forge paradigm, allowing general-purpose and forged artefacts to coexist as service provision progressively shifts towards forged artefacts that pass verification. Over the longer term, demand-specific production may shift value accumulation towards production systems and extend trust from individual artefacts to the systems that produce them.

\backmatter

\bibliography{references}

\begin{appendices}
\counterwithin*{table}{section}
\renewcommand{\theHtable}{\thesection.\arabic{table}}

\section{Mechanism analysis and controlled comparisons}
\label{app:ablations}

\subsection{Deep specialisation and vertical integration}

Section~\ref{sec:evaluation} reports the final performance of the three Forge systems. This section explains how those gains arise from reorganising implementations around specific demands rather than simply adding more runtime scheduling.

In ForgeTrain, production reorganises training execution around a fixed model, tensor shapes and parallel layout. A general training framework typically executes gradient clipping, the Adam update and precision conversion as separate operations, requiring multiple parameter passes and kernel launches. ForgeTrain fuses these operations into one kernel so that gradient processing, parameter updates and write-back share a data-access path. It also organises a microbatch's forward pass, loss computation and backward pass as one CUDA Graph replay, reducing per-layer dispatch and host involvement.

This integration depends on fixed task conditions. Optimiser state, clipping coefficients and parameter layouts can be determined for the target model, moving some decisions that would otherwise occur at runtime into construction. ForgeTrain's optimisation trajectory shows that kernel fusion, communication--computation overlap, device-side gradient-norm computation and zero-copy gradient views jointly improve MFU\cite{forgetrain2026}.

ForgeMegakernel follows a related production strategy. It arranges GPU computation, data loading, on-chip storage and synchronisation for decoding within one large kernel, adjusting thread organisation and intermediate layouts for the target model, batch size and sequence length. Compared with launching multiple general-purpose kernels, this organisation reduces intermediate write-back, launch and synchronisation overhead. The gain comes from coordinating components across boundaries\cite{forgemegakernel2026}.

These cases show that the production side translates known demand conditions into implementation constraints and uses them to redraw component boundaries. Specialisation determines which general mechanisms can be removed, while vertical integration determines which previously separate operations can be combined; together they enlarge the performance space available to specialised artefacts.

\subsection{Independent judgement and quality constraints}

All three Forge systems separate artefact production from result judgement. The production side may change implementations, search routes and resource allocation; the verification side uses independent references, fixed checkers or predefined quality requirements to decide whether a candidate is acceptable. This prevents the production side from creating apparent performance gains by changing the conditions of evaluation.

ForgeTrain's comparisons (Table~\ref{tab:harness_eval}) expose two problems caused by missing independent quality constraints. After the complete production and checking organisation is removed, the agent lowers optimiser precision from FP32 to BF16 to increase throughput. The modified engine may run faster, but the change alters the optimiser's numerical computation and does not establish training-quality parity, so it cannot count as a performance gain satisfying the quality requirements.

A second comparison retains the construction organisation but removes the explicit training-quality requirement. On MiniCPM4 0.5B, MFU remains around 37\% within 33 rounds, whereas the complete process reaches 46.1\%. The trajectory shows that, without a clear quality criterion, the agent treats bitwise agreement as an immutable restriction and avoids changes to reduction order and operator implementations. An explicit training-quality requirement lets it move from bitwise agreement to training-quality parity, expanding the acceptable optimisation space while preserving the final judgement condition\cite{forgetrain2026}.

The objects of judgement differ across the three settings. The training framework requires checks of long-run trajectories, final loss and downstream model quality; decoding kernels require checks against an independent numerical reference and of service-level task quality; computational kernels must pass the official or benchmark tolerance test associated with their operator source. Local checks screen candidates, integration checks examine component interactions and comprehensive evaluations establish delivery readiness. These checks jointly form the acceptance evidence, while a local pass cannot replace final judgement.

ForgeKernel's operator comparisons also show why judgement must be independent of the implementation being optimised. Each candidate must pass the official or benchmark tolerance test associated with its operator source before performance measurement, and the harness rejects candidates that call the reference implementation directly or delegate computation to it\cite{forgekernel2026}.

Quality constraints therefore serve two roles: they prevent unverified numerical or behavioural changes from entering delivered artefacts, and they tell production which changes are acceptable. A performance result becomes an acceptance result only when paired with the corresponding quality evidence.

\begin{table}[!hbp]
\small
\caption{Mechanism comparisons in the Forge systems.}\label{tab:harness_eval}
\begin{tabular*}{\textwidth}{@{\extracolsep\fill}L{2.4cm}L{3.0cm}L{3.4cm}L{3.1cm}@{}}
\toprule
Setting & Change or comparator & Reduced configuration & Complete system \\
\midrule
Training framework & Remove harness (MiniCPM5 16B-A3B) & Lower optimiser precision; no verified quality-parity gain & MFU 27.6\%; reference 23.5\% \\
Training framework & Remove quality requirement (MiniCPM4 0.5B) & MFU about 37\% within 33 rounds & MFU 46.1\% \\
\botrule
\end{tabular*}
\end{table}

\section{Production, judgement and reuse in other systems}
\label{app:other_systems}

The preceding sections report performance and quality evaluations of the three Forge systems, together with comparisons involving forged-artefact feedback and quality requirements. Other systems also produce artefacts around concrete demands, evaluate candidates with execution checks and adjust later construction from feedback. This section analyses these methods through production, judgement, feedback and reuse (Table~\ref{tab:evidence}).

\subsection{Demand-specific specialisation and integration}

Bespoke OLAP constructs a complete analytical engine around a specified query workload. It arranges storage layouts and data structures from query patterns, generates ingestion and query-execution code, and then optimises the target queries. Because the workload is fixed, the system can specialise predicates, data layouts and execution paths without retaining every abstraction needed by a general database. Its single-threaded TPC-H and CEB results reach 11.17 and 45.33 times DuckDB, respectively\cite{bespokeolap2026}.

Sol divides video-inference optimisation into local tasks for caching, sparse attention, pruning, quantisation and kernel fusion. Different agents explore candidates in parallel, after which an integration agent constructs a complete acceleration stack. Integration must handle accumulated approximation error, interface compatibility and resource contention. Human validators check video quality on a fixed validation set, and their feedback determines whether later optimisation becomes more or less aggressive. The system shows that a local technique has delivery value only when the integrated artefact still satisfies quality requirements\cite{sol2026}.

These systems embody the same principle: the production object is a complete artefact jointly designed around a demand, rather than an isolated code fragment. The more explicit the demand conditions, the more the production side can remove general abstractions and reorganise execution across component boundaries.

\subsection{Process management and artefact judgement}

CAKE separates single-shape kernel generation from kernel-family integration. For one input shape, it applies compiler-legality checks, numerical verification and performance measurement; after forming a high-performance kernel, it organises the kernel into an implementation and dispatcher covering multiple shapes. Integration then checks held-out inputs, boundary shapes, dispatch conditions and fallback paths. Erroneous or slow candidates return for further modification\cite{cake2026}.

Bespoke OLAP first checks query results against DuckDB before timing and tracing; regression checks continue during optimisation, and erroneous modifications are reverted. Sol similarly checks overall visual quality after local acceleration techniques have been integrated into a complete video-inference stack. Passing a local implementation therefore does not establish that the complete artefact is ready for delivery\cite{bespokeolap2026,sol2026}.

\subsection{Execution feedback, search and experience reuse}

K-Search organises parent programs and optimisation intentions as search actions. It selects actions by priority and evaluates local modifications through code generation and execution; after several unsuccessful attempts, it ends the current direction and uses the history to add, adjust or prune search branches\cite{ksearch2026}. The search state records both the current program and the tested optimisation hypotheses, allowing later search to continue from existing programs and feedback.

The skill-based harness uses representative-workload screening followed by full-workload acceptance. Candidates first undergo inexpensive tests on representative workloads, and only promising candidates enter full evaluation; promotion requires overall improvement without a correctness regression. Successful and failed records enter an archive, and humans adjust the direction when search stalls\cite{harnesskernel2026}.

These systems illustrate two forms of reuse. K-Search reuses search states and optimisation intentions, while the skill-based harness reuses artefacts that have passed checks together with their test records. The former reduces repeated trials in later search; the latter lets later construction start from a reliable effective state.

Across these systems, production, judgement and feedback are connected rather than independent steps. Production generates candidates, judgement supplies traceable execution evidence and feedback adjusts later production. Once artefacts and search experience are retained, later construction can continue from an existing state. This corresponds to forged-artefact feedback, knowledge reuse and independent judgement in ForgeSystem.

For Table~\ref{tab:evidence}, the correspondences are our synthesis of the source methods; performance results use the evaluation settings of each source.

\begin{table}[t]
\footnotesize
\caption{Production, judgement and reuse mechanisms in other systems.}\label{tab:evidence}
\begin{tabular*}{\textwidth}{@{\extracolsep\fill}L{2.0cm}L{3.6cm}L{2.5cm}L{3.0cm}@{}}
\toprule
System & Production and judgement mechanism & Feedback or reuse & Representative result \\
\midrule
Bespoke OLAP\cite{bespokeolap2026} & Storage layout, query execution and parallelisation specialised to workload; query checks and regression rollback & Continue optimisation from checked engine; revert failed modifications & TPC-H / CEB single-thread speedup over DuckDB: 11.17 / 45.33$\times$ \\
CAKE\cite{cake2026} & Single-shape kernel and kernel-family construction; checks on dispatch, held-out inputs and fallback paths & Return unsuitable candidates for further modification & B200, six BF16 KDA-prefill shapes: 2.05$\times$ geometric-mean speedup over FlashKDA \\
Sol\cite{sol2026} & Parallel local optimisation, full-stack integration and human quality checks & Use overall quality feedback to adjust later optimisation & Three video models: 2.27, 2.38 and 2.77$\times$ over SGLang \\
K-Search\cite{ksearch2026} & Search tree of optimisation intentions; execution results update priority and pruning & Continue from existing programs and history & H100 TriMul official evaluator, local test: 1030 microseconds geometric-mean latency \\
Skill-based harness\cite{harnesskernel2026} & Representative-workload screening, full-workload acceptance and candidate promotion & Retain successful and failed records; recover direction when stalled & Five B200 contest operator classes: 1.12--29.68$\times$ over supplied FlashInfer references \\
\botrule
\end{tabular*}
\end{table}

\section{Scope and bottlenecks}
\label{app:scope}

The scope and bottlenecks of Forge centre on three factors.

\textbf{Specialisation benefits.} Rebuilding offers limited value when general-purpose solutions already satisfy a demand or demand differences yield little performance or efficiency gain. Specialisation must also preserve necessary interoperability conventions.

\textbf{Complete production cost.} Cheaper model calls do not imply proportionately cheaper delivery. Experimentation, verification, integration and human intervention may remain costly, particularly where reliable evaluation criteria are unavailable.

\textbf{Returns from accumulated experience.} Prior work shows that construction experience can improve model performance\cite{swegym2025,alphaproof2026}, but repetitive trajectories and limited learning signals may lead to diminishing returns. Sustained capability gains from the outer loop require validation over multiple rounds of construction and training.

\end{appendices}

\end{document}